\documentclass[a4paper,12pt]{report}

\usepackage{amsmath}
\usepackage{comment}
[
        a4paper,
        left=3cm,
        right=3cm,
        top=1.5cm,
        bottom=1.5cm,
]
\usepackage{amsmath,amssymb}
\usepackage{amsmath}
\usepackage{textcomp}
\usepackage{wasysym}
\usepackage{graphicx}
\usepackage{xcolor}
\usepackage{caption}
\usepackage{subcaption}
\usepackage{float}
\usepackage{cite}
\usepackage{hyperref}
\usepackage{lipsum}
\usepackage{breqn}
\usepackage{siunitx}
\usepackage{pdfpages}

\begin{document}
\titlepage
\vspace{1.5cm}
\begin{center}
{\large \bf {Network Model of Active Fluctuations of Thin Elastic Shells Swollen by Hydrostatic Pressure}}\\

\vspace{0.5cm}
Ajoy Maji 
and
Yitzhak Rabin\\
Department of Physics and Institute of Nanotechnology and Advanced Materials, Bar-Ilan University, Ramat-Gan 5290002, Israel
\end{center}

\begin{abstract}
    Many organisms have an elastic skeleton that consists of a closed shell of epithelial cells that is filled with fluid, and can actively regulate both elastic forces in the shell and hydrostatic pressure inside it. In this work we introduce a simple network model of such pressure-stabilized active elastic shells in which cross-links are represented by material points connected by non-linear springs of some given equilibrium lengths and spring constants. We mimic active contractile forces in the system by changing the parameters of randomly chosen springs and use computer simulations to study the resulting local and global deformation dynamics of the network. We elucidate the statistical properties of these deformations by computing the corresponding distributions and correlation functions. We show that pressure-induced stretching of the network introduces coupling between its local and global behavior: while the network opposes the contraction of each excited spring and affects the amplitude and relaxation time of its deformation, random local excitations give rise to contraction of the network and to fluctuations of its surface area.
\end{abstract}

\section{Introduction}
 The best known example of a thin elastic shell filled with fluid and stabilized by the interplay of hydrostatic pressure and elastic forces is a swollen baloon \cite{1_kier2012diversity}. In nature, a diverse group of organisms have an elastic skeleton that often consists of epithelial cells, and is filled with fluid\cite{2_mosaliganti2019size}. Unlike baloons, living systems can regulate both elastic forces and hydrostatic/hydraulic pressure by contraction of supracellular actomyosin fibers\cite{3_balasubramaniam2021investigating}, and by osmotic control of the hydrostatic pressure inside the shell\cite{4_kucken2008osmoregulatory}, respectively.  A simple example is hydra, a small fresh water organism in which the elastic shell consists of two layers of epithelial cells filled with water. During the process of its regeneration the tissue forms a closed spherical shell which eventually develops into an elongated shape of the mature organism (the process can be reversed by the application of electric fields)\cite{5_braun2019electric,braun2021calcium}.
Contraction of the supra-cellular actomyosin fibers that run through the epithelial cells (an active process that involves ATP hydrolysis) leads to deformations and morphology changes of the sphere \cite{6_livshits2017structural}. Multi-cellular spheroids are another example of an elastic hydraulic shell whose morphology can be affected by physical fields (radial flow, electric current)\cite{7_delarue2014stress,8_duclut2021hydraulic}, a problem that has been theoretically studied using methods based on the active gel model\cite{9_kruse2005generic,10_prost2015active}. Such continuum viscoelastic models have been applied to a broad range of active phenomena in bulk systems\cite{11_marchetti2013hydrodynamics,12_streichan2018global}, active surfaces and sheets\cite{13_salbreux2017mechanics, Bernheim2018}, and active elastic shells, e.g., the role of the actomyosin cortex in cell division\cite{14_turlier2014furrow} and its effect on the morphology of growing cell aggregates. Note that while continuum models of active viscoelastic systems contain the solid as a limiting case, they have been mostly applied to viscous-stress-dominated long-time phenomena and the analysis was limited to the fluid case\cite{8_duclut2021hydraulic,14_turlier2014furrow,15_metselaar2019topology,16_tlili2020migrating}. Nevertheless, models of active elastic solids (cross-linked gels) have also been studied both analytically and numerically\cite{17_banerjee2011generic,18_kopf2013non}.

In this paper we explore an alternative approach to the study of pressure-stabilized active elastic shells:
rather than use a continuum description of such closed surfaces, we introduce a simple network model in which
cross-links are represented by material points connected by non-linear springs of some given equilibrium lengths and spring constants. Following the
methodology we developed to describe phase transitions in two-dimensional sheets of cross-linked polymers\cite{19_peleg2007filamentous}, in addition to the spring
forces we introduce body forces acting on each cross-link, that represent pressure differences between the
inside and the outside of the closed shell. This pressure can be tuned, e.g., by changing the external osmotic
pressure on the shell or by active control of transport of osmolytes through it. The balance between the
elastic and the hydrostatic forces acting on the cross-links guarantees a stable mechanical equilibrium state
of the network. Analogously to our work on active phase separation in multi-component systems where we
introduced active forces by changing the interactions of selected particles with their environment (e.g., turning
attraction into repulsion)\cite{20_osmanovic2019chemically}, in the present model we implement active contractile forces (excitations of springs) by changing the equilibrium
lengths and the spring constants of randomly chosen springs, at regular or at randomly chosen time intervals. This results in
contraction of the excited springs and, because of global connectivity constraints (the topology of the network
is fixed, reflecting the solid character of our system), the effects of such contractions propagate through
the entire network and result in its global deformation. We use molecular dynamics (MD) to study both the
local and the global response of the network to such local excitations. 
Note that unlike vertex models of epithelial tissues where each cell is represented as a polygon with vertices
and edges are shared between adjacent cells \cite{21_fletcher2014vertex,22_barton2017active, 23_perez2020vertex}, our network model does not attempt at a cell-level
description of biological systems and therefore does not capture phenomena such as cell growth,
division and death. The advantages of our model are its simplicity and computational speed (a typical
computation of statistical distributions takes about 5 minutes on a personal computer), that make extensive
exploration of the parameter space feasible and allow us to study and to visualize morphology changes
and network fluctuations in three dimensional space and in time.

In the Model section we introduce our network model of a thin elastic shell. We use triangulation of the surface of a fullerene molecule ($C_{60}$) to create the ground state (in the absence of hydrostatic pressure) configuration of the network in which the vertices of the triangles correspond to cross-links and the sides represent elastic springs. This choice is motivated by simplicity considerations (this is a straightforward method to triangulate a quasi-spherical surface) but comes at the price of having four different classes of equilibrium spring lengths and three different types of triangles (see Fig. S1 in SI). Nevertheless, as we will argue in the following, the emerging physical picture does not depend on the details of network structure. We define the hydrostatic,  elastic and friction forces that act on each vertex, write down the equations of motion and introduce the algorithm used to solve them numerically.  In the results section we report the results of computer simulations of our model in the presence of active contractile excitations generated by randomly choosing network springs and changing their equilibrium lengths and spring constants. We study the statistical properties of both the global (total surface area of the system) and local (lengths of springs and areas of triangles) properties of the system: we calculate both their steady state distributions (by averaging over the network and over time) and their temporal correlations (by computing the corresponding two-time auto-correlation functions). 
Next, we study the effects of direct coupling between the total area of the system and the hydrostatic pressure, both in the case of positive (pressure increases with increasing area) and of negative (pressure decreases with increasing area) feedback. Finally, in the discussion section we summarize our results, outline connections with recent experiments on hydra regeneration. Finally we discuss the limitations and possible extensions of our work.

\section{Model} 
\subsection{Construction of the network}

\begin{figure}[t!]
        \centering
        \includegraphics[width=0.9\linewidth]{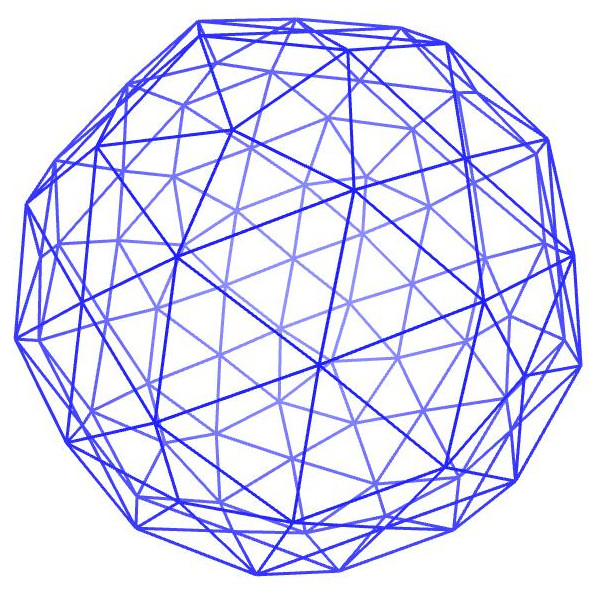}
        \caption{Triangulation of a fullerene. Sides of triangles on the near (far) side of the fullerene  are shown in dark (light) blue.}
        \label{model}
\end{figure}

 To create a closed two-dimensional network on a quasi-spherical surface, we carried out a triangulation of a symmetric fullerene (e.g., $C_{60}$ molecule) that consists of 20 hexagons and 12 pentagons. We connected the centroid of each polygon to the vertices of that polygon such that each pentagon (hexagon) contains five (six) triangles. The resulting network is made of 180 isosceles triangles that can be classified into three groups, each of different area (see Fig.\ref{model}). In the following we will treat the $92$ vertices of the triangles as cross-links and replace the $270$ sides of the triangles by elastic springs of equilibrium lengths determined by the geometry of the fullerene. In choosing this particular topology of the network we were guided by considerations of symmetry and convenience (the fullerene molecule has a quasi-spherical shape and its geometric parameters are well-known). Clearly, other network topologies are possible as well, in particular, that of a random network constructed by Voronoy tessellation of a sphere \cite{32_augenbaum1985construction, chen2003algorithm}. We plan to implement such constructions in future work in order to explore the effects of network topology on its dynamics.

\subsection{Equations of motion:} 

The area vector of triangle $i$ can be represented as $\Vec{a_{i}}=a_{i} \Hat{n_{i}}$, where $\Hat{n_{i}}$ is the unit normal and $a_{i}$ is the area of the triangle. Introducing the hydrostatic pressure $p$ that swells the closed shell, we define the hydrostatic force $\Vec{F^h_{k}}$ on vertex $k$  as the vector sum of the contributions of the  triangles that meet at this vertex:
    \begin{equation}
    \Vec{F^h_{k}}=p\sum_{i=1}^{m} a_{i} \Hat{n_{i}}, 
    \end{equation}
where $m=6$ ($m=5$) if the vertex is located at the center of a hexagon (pentagon).    
The total area (A) of the system is given by    
\begin{equation}
    \Vec{A}=\sum_{i=1}^{n}\Vec{a}_{i}
\end{equation}

For linear elastic springs the force on  vertices $k$ and $q$ connected by a spring, is given by
\begin{equation}
    \Vec{F}^s_{kq}=-K\frac{\Vec{r_k}-\Vec{r_q}}{|\Vec{r_k}-\Vec{r_q}|}(|\Vec{r_k}-\Vec{r_q}|-l^{eq}_{kq})
\end{equation}

where $\Vec{F}^s_{kq}$ represents spring force on vertex $k$ due to the spring connecting it to vertex $q$, (clearly, $\Vec{F}^s_{qk}=-\Vec{F}^s_{kq}$), $K$ is the spring constant, and $l^{eq}_{kq}$ is the equilibrium length of this spring which we take to be the length of the $k-q$ bond of the fullerene.
      
     The total elastic force on vertex $k$ is the sum of the contributions of all springs that connect this vertex to its immediate neighbors: 
      \begin{equation}
    \vec{F}^s_{k}=\sum_{q=1}^{m} \Vec{F}^s_{kq}
    \end{equation}
      In addition to hydrostatic and spring forces there is a frictional force acting on each moving vertex which is proportional to its instantaneous velocity $ -\zeta \vec{v}_{k}(t)$, where $\zeta$ is the friction coefficient.
    Hence the equation of motion for the $k^{th}$ vertex is,
    \begin{equation}
        \dot{\Vec{v}}_{k}(t)=-\zeta \Vec{v}_{k}(t)+\Vec{F}_{k}(t)
        \label{equ_motion}
\end{equation}

where $\Vec{F}_{k}=\Vec{F}^h_{k}+\Vec{F}^s_{k}$. Unless otherwise specified,  we set the mass of each vertex to unity and its friction coefficient to $6$ and therefore the damping time is $1/6$ in time units used throughout our simulation. In our simulations we take the MD time step to be $\Delta=0.01$.
 
 Note that the velocities of all vertices must vanish in steady state. In the absence of hydrostatic pressure ($p=0$) and other external forces (e.g., active excitations of the springs), this implies that the spring forces on each vertex $k$ must vanish as well, $\Vec{F}^s_{k}=0$. This defines the ground state of our system which, in the following, we will refer to as the equilibrium state (since thermal fluctuations are neglected, this corresponds to mechanical equilibrium in the absence of external forces on the network). The equilibrium lengths of all springs are given by the lengths of the corresponding sides of the triangles in the triangulated fullerene.

 The equations of motion were solved using the following algorithm for simultaneously updating the velocity and the position of each vertex $k$ (see derivation in Supplementary Information):

 \begin{equation}
         \Vec{v}_{k}\big((n+\frac{1}{2})\Delta)=e^{-\zeta \Delta}\Vec{v}_{k}\big((n-\frac{1}{2})\Delta)+e^{-\zeta \frac{\Delta}{2}} \hspace{0.1cm} \Delta \hspace{0.1cm}\Vec{F}_{k}(\Vec{r}_{k}(n \Delta))
         \label{velocity_eq}
     \end{equation}
and
\begin{equation}
    \Vec{r}_k((n+1)\Delta)=\Vec{r}_k(n\Delta)+\Delta \hspace{0.1cm}\Vec{v}_k \hspace{0.1cm}((n+\frac{1}{2})\Delta)
    \label{position_eq}
\end{equation}
where $n$ is the number of MD time steps. Thus, in order to compute the velocities of the vertices at time $t$ one has to specify their velocities at time $t-\Delta$ and the forces acting on them at time $t-\Delta/2$; the positions at time $t$ are computed from the positions at time $t-\Delta$ and the velocities at time $t-\Delta/2$.

\subsection{Nonlinear springs:}
Initially, we used Hookean springs in our simulation (the restoring force of a spring is linearly proportional to its elongation). We found that linear elasticity is not sufficient to maintain the force balance in the system and to reach a steady state. As the system inflates under the action of hydrostatic pressure $p$, the areas of the triangles increase and since the hydrostatic force on a triangle is given by the product of the pressure and the area, the hydrostatic force on each vertex increases with swelling of the network. Even though the spring forces increase as well, their projection on the radial direction (in opposite direction to the hydrostatic force), decreases. The restoring forces produced by the linear springs can not counterbalance the hydrostatic forces and this results runaway expansion of the system.
 
In view of the above, we decided to use non-linear elastic springs in our simulation (note that non-linear elastic behavior is often observed in biopolymer networks \cite{Burla2019}). We introduced an elongation-dependent spring constant:
    \begin{equation}
    K_{pq}=K\left\{[\frac{|\Vec{r_p}-\Vec{r_q}|}{l^{eq}_{pq}}-1]^2\right\}
    \end{equation}
    where $l^{eq}_{pq}$ is the equilibrium length of the spring that connects the vertices $p$ and $q$. The instantaneous elastic force is then given by
    \begin{equation}
         \Vec{F}^s_{kq}=-Kl^{eq}_{kq}\frac{\Vec{r_k}-\Vec{r_q}}{|\Vec{r_k}-\Vec{r_q}|}[\frac{|\Vec{r_k}-\Vec{r_q}|}{l^{eq}_{kq}}-1]^3
\label{nonlinear force}
\end{equation}

\section{Results}
\subsection{Active Excitations}
\subsubsection{Single excitation}
Consider a network in steady state in presence of hydrostatic pressure $p_0=0.1$. This steady state is prepared by starting from the ground state configuration at $p=0$, raising the pressure to $p_0=0.1$ and allowing it to relax to the new steady state (see Fig. S2 in SI). Next, we randomly choose one of the springs in the network  and excite it by reducing its equilibrium length from $l^{eq}$ to $l^{eq}/2.0$ and increasing its spring constant from $K=1$ to $K_{ex}=10$ (note that this increases the amplitude of the elastic force from $Kl^{eq}$ to $5Kl^{eq}$ in Eq. \ref{nonlinear force}).   We refer to it as active excitation since it involves changing the parameters of the corresponding spring by some external means. 
This leads to contraction of the excited spring and ultimately to the deformation of the entire network. The parameters are chosen to make the excitation of a single spring to be sufficiently strong to produce a readily observable deformation of the network.

\begin{figure}[t!]
        \centering
        \includegraphics[width=1.0\linewidth]{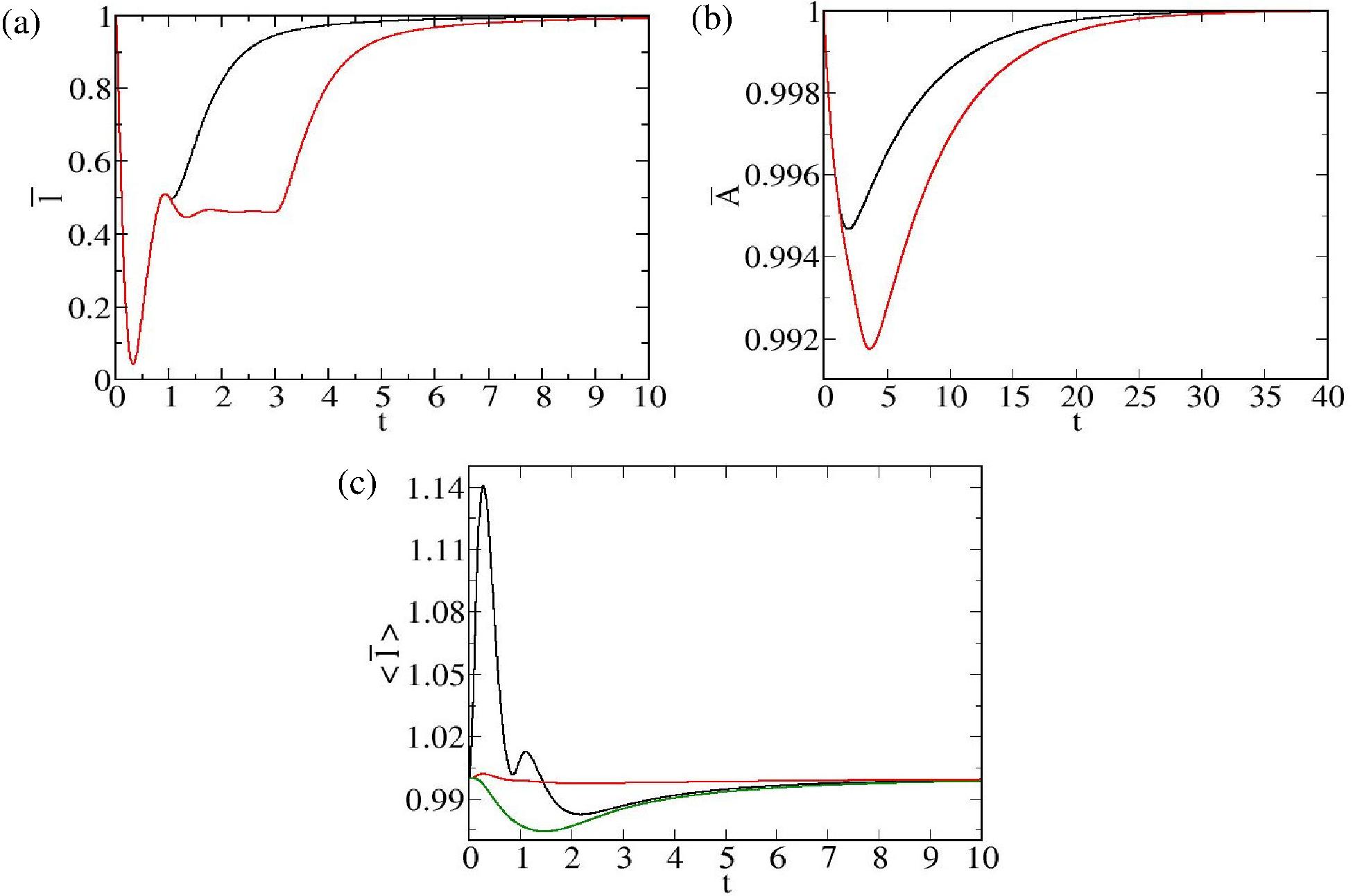}
        \caption{Response to a single excitation at $t=0$: (a) Normalized length of the excited spring as a function of time. The black and red curves represent relaxation times $t_{r}=1$ and $t_{r}=3$, respectively. (b) Normalized total surface area of the system is plotted as a function of time (color code is the same as in (a)). Note the different time scales in (a) and (b). (c) Average normalized length of nearest neighbor springs (in black), and of next nearest neighbor springs (in green), is plotted as a function of time. The  average normalized length of all the bonds in the network is shown in red.}
        \label{single_excitation_length}
\end{figure}

In Fig. \ref{single_excitation_length}(a) we plot the normalized length of the excited spring $\bar{l}=l/l_s$, where $l_s$ is length of the spring in steady state with hydrostatic pressure $p=0.1$ (in the absence of excitations). We assume that the spring is excited at time $t=0$ and that the excitation relaxes instantaneously (the spring constant and the equilibrium length return to their unperturbed values, $K=1$ and $l^{eq}$) at time $t_r$. The curves corresponding to $t_r=1$ and $t_r=3$ are shown in the figure. Following the change of parameters, the length of the excited spring rapidly (in about $0.3$ time units) drops to below $5\%$ of it's steady state value and then recoils to about $50\%$ of this value in about $1$ time unit (see the shoulder at $t=1$, for both values of $t_r$). At longer times, the relaxation to steady state depends on the value of $t_r$: for $t_r=1$ the length of the spring increases smoothly with time and recovers more than $90\%$ of its steady state length in about $3$ time units after the onset of excitation. For $t_r=3$, the length of the excited spring first stabilizes at a value slightly below $l_s/2$, a point at which the contractile force on the excited spring is balanced by the elastic forces applied on it's ends by the rest of the stretched network (this corresponds to a steady state of a network that is stretched by hydrostatic forces and contains one excited spring). Then, at $t=3$ the excitation relaxes, the spring recovers it's original parameters and is driven by the elastic forces in the network reaches to it's unperturbed steady state length $l_s$ (it recovers more than $90\%$ of $l_s$ in about $6$ time units from the beginning of the excitation).
The response of the entire network to the deformation produced by a single excited spring, can be monitored by following the time series of the normalized total area of the network, $\bar A=A/A_s$ ($A_s$ is the steady state area in the absence of excitations), shown in Fig. \ref{single_excitation_length}(b). Note that the relaxation of the total area is much slower than that of the excited spring (compare Figs. \ref{single_excitation_length}(a) and (b)). 

All the above results about the local and the global deformations of the network were obtained for $p_0=0.1$. In order to see how the deformations depend on hydrostatic pressure, in Figs. S3(a) and (b) we compare the time traces of $\bar l$ and $\bar A$ with time, following the excitation of a single spring in a network in steady state for different values of $p_0$. We find that increasing pressure results in stronger contraction (with respect to the steady state values $l_s$ and $A_s$, both of which increase with pressure) and in faster relaxation to steady state, of both $\bar l$ and $\bar A$. 

We proceed to examine the dependence of the local and global response of the network to an isolated excitation, on the various parameters in the problem. We find that the depth of the minimum of the normalized spring length $\bar l$ exhibits strong dependence on the friction coefficient $\zeta$, the spring constant $K$ and mass $m$: it decreases with increasing $\zeta$ (not shown) and with increasing $K$ and $m$ (see Fig. 4(a) in SI). These trends are clearly related to the forces that oppose the contraction of the excited spring: the frictional force ($\propto \zeta$), the elastic force exerted on the spring by the rest of the network ($\propto K$) and the inertia of the network ($\propto m$). Interestingly, the shoulder at $t\approx 1$ disappears and the relaxation back to steady state becomes slower when the mass of the vertices is increased from $m=1$ to $m=2$. Similar trends are observed in the time series of the normalized total area $\bar A$ shown in Fig. S4(b) in SI. The depth of the minimum in $\bar A$ decreases with increasing $K$ and $m$ but they have different effects on the relaxation time of the total area back to steady state - relaxation becomes faster with increasing $K$ (stronger restoring forces of the network) and slows down with increasing $m$ (because of increased inertia of the network).

How does the perturbation introduced by the excited spring propagate through the network? In order to answer this question we first identified the nearest neighbors (nn) and next nearest neighbors (nnn) of the excited spring. The nn springs are springs connected to either of the two ends of the excited spring and the nnn springs are connected to the ends of the nn springs, but not directly to the excited spring. We then monitor how the average lengths of nn and nnn springs change with time following the excitation. Since during the contractile excitation, the length of the excited spring is shortened, we expect that the immediate (before the deformation can propagate to the rest of the network) reaction of the nn springs are to increase their length.

Simultaneously with the rapid contraction of the excited spring, a sharp spike in the average length $<\bar l>_{nn}$ of the nearest neighbor springs (averaged over the normalized lengths of all nearest neighbor springs) is observed in Fig.~\ref{single_excitation_length}(c) (black curve). The spike is followed by oscillatory decay and eventually relaxation to the steady state value. The nnn springs initially respond to the excitation by shortening their length $<\bar l>_{nnn}$ (averaged over the normalized lengths of all next nearest neighbor springs); at longer times this average length relaxes to its steady state value (green curve in Fig. \ref{single_excitation_length}(c)). As expected, the intensity of the perturbation decreases with distance from the excited spring. Only a minor perturbation (a small peak, followed by decay to steady state) of the average length $<\bar l>_{all}$ of all network  springs is observed in Fig.~\ref{single_excitation_length}(c) (red curve).

\subsubsection{Multiple periodic and non-periodic excitations:}

We begin by preparing a network in its ground state (with $p=0$) and let it expand by introducing hydrostatic pressure $p=0.1$. After the system reaches a steady state, we begin to introduce active excitations by randomly choosing a spring and  changing its equilibrium length from $l^{eq}$ to $l^{eq}/2.0$ and its spring constant from $K=1$ to $K=10$. The parameters of each excited spring relax back to their equilibrium values ($l^{eq}$ and $K=1$) after time $t_{r}=1$. Note that the excited spring does not return immediately to its steady state length; in fact it takes additional $2$ time units to regain more than $90\%$ of its steady state length (see Fig. \ref{single_excitation_length}(a)). 

In the case of periodic excitations, we randomly choose a spring and excite it. Then, after time $\delta t$ we randomly choose another spring, excite it and repeat the process indefinitely. In the following we choose a fixed time interval $\delta t =3$ between excitations. Since this time interval is larger than the spring relaxation time $t_r=1$, each excited spring relaxes almost completely to its steady state length by the time next excitation occurs in the network. Note however, that since the total area of the network takes a much longer time to relax (see Fig. \ref{single_excitation_length}(b)), one expects the area fluctuations to show deviations from a purely periodic profile, even in the case of perfectly periodic excitations of individual springs, as indeed observed in Fig.~\ref{without_feedback}(a).

In case of non-periodic excitations, the time interval between subsequent excitations is a random number taken from a uniform distribution in the range $[0-2]$, with an average $<\delta t>=1$. The excitation parameters are the same as in the case of periodic excitations described above: the equilibrium length changes from $l^{eq}$ to $l^{eq}/2.0$ and the spring constant from $K=1$ to $K=10$. Each excited spring switches back to its ground state parameters exactly one time unit after the onset of its excitation ($t_{rel}=1$). Note that because the interval between excitations varies between $0$ and $2$ time unit, the number of excitations (i.e., the number of springs with changed equilibrium length and spring constants) that exist in the network at any instant of time,  varies between $0$ and $5$ (the distribution of the number of simultaneous excitations in the network is shown in Fig. S5). A movie of a non-periodically excited network is shown in Supplementary Information (see movie M1 in SI).  

\subsubsection{Global dynamics: Total surface area :}

\begin{figure}[t!]
        \centering
        \includegraphics[width=1.0\linewidth]{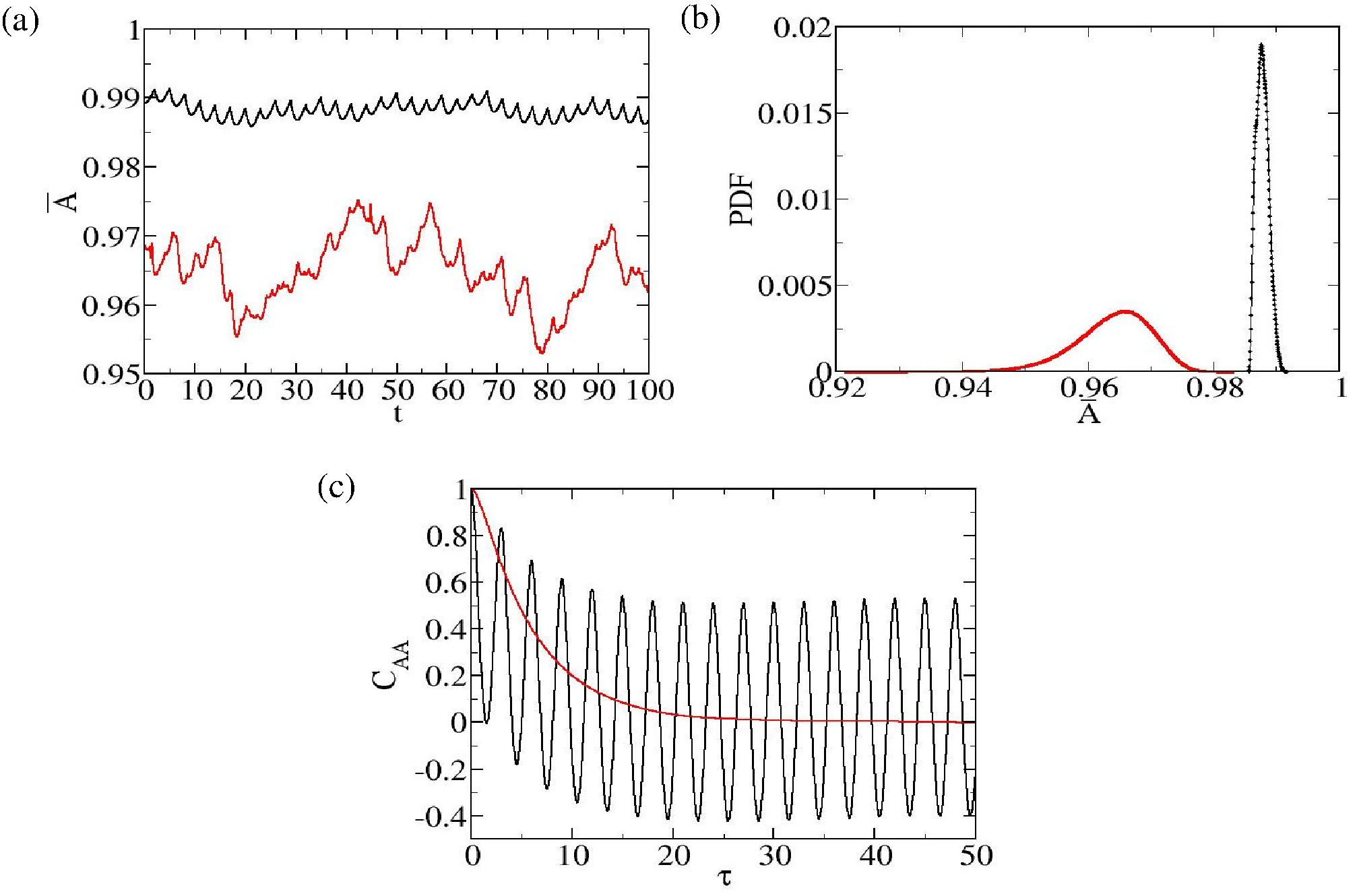}
        \caption{Comparison of (a) time series, (b) probability distribution functions, (c) auto-correlation functions of the normalized total surface area $\bar{A}$, for periodic (black curves) and non-periodic (red curves) excitations. }
        \label{without_feedback}
\end{figure}

We begin by considering the effect of excitations on the global properties of the system. In
Fig. \ref{without_feedback}(a) we plot the normalized total surface area $\bar{A}=A/A_s$ ($A_s$ is the surface area in steady state with hydrostatic pressure $p_0=0.1$, without excitations) as a function of time for periodic and  non-periodic excitations. It is evident from Fig.~\ref{without_feedback}(a) that the statistical properties of the fluctuations of the total area are different in the two cases. In particular, both the deviation of the mean of $\bar A$ from the steady state value $\bar A=1$, and the amplitude of fluctuations around the mean  (i.e., the variance of $\bar A(t)$) are larger for non-periodic than for periodic excitations. We also observe that the
fast fluctuations  are sitting on a slowly varying envelope (this effect is more pronounced in the non-periodic case), suggesting the existence of multiple time scales in the system .

Next, we computed the probability distribution function (PDF) of the total area by treating the values of $\bar{A}$ measured at different times as a statistical ensemble, and constructing a histogram of the number of times a particular value of the area $\bar{A}$ occurs during the measurement time ($9\cdot10^7$ MD time steps or $9\cdot10^5$ time units). In order to get a PDF, we normalized the area under the resulting curve to unity. Although each such time series depends on the particular realization of the dynamics of the system (the order in which the different $270$ springs are excited is determined by a random number generator, and in the case of non-periodic excitations, the time of excitation is a random number in the range from $0$ to $2$ time units), different realizations of the time series of the total area yielded identical PDFs.

The information obtained from visual inspection of the time series of the total area in Fig. \ref{without_feedback}(a), is summarized by the corresponding probability density functions (PDFs) shown in Fig. \ref{without_feedback}(b). For periodic excitations (black curve) the distribution is centered at a larger value of the area and is much narrower than for non-periodic ones (red curve).  The PDF for non-periodic excitations has a long tail towards lower values of $\bar A$ whereas the distribution for periodic excitations has a tail towards higher values of $\bar A$. In order to understand this behavior we notice that while there exists at most one excitation in the network in the periodic case (recall that the time between excitations is larger that their relaxation time), the number of excitations that can be simultaneously present in the system varies between $0$ to $5$ in the non-periodic case (see Fig. S5 in SI); recall that since excitations are always contractile, their direct effect is to decrease the total area $A$. The weak tail towards higher A values in the periodic case, is due to the contribution of time intervals in which the length of an excited spring has nearly relaxed to its steady state value before the onset of a new excitation (see Fig. \ref{single_excitation_length}(a)). As the length of this spring increases, the network begins to relax towards its steady state configuration, $\bar A=1$, giving rise to the observed tail (black curve in Fig. \ref{without_feedback}(b)).

In order to get additional information about the global dynamics of the system, we computed the auto-correlation function of the total surface area A,
\begin{center}
\begin{equation}
    C_{AA}(\tau)=\frac{<\bar{A}(t+\tau)\bar{A}(t)>-<\bar{A}(t)>^2}{<[\bar{A}(t)]^2>-<\bar{A}(t)>^2}
    \label{pearson_c}
\end{equation}
\end{center}

where, for each value of the delay time $\tau$, the averaging is taken over the time $t$ (in the interval from $1$ to $9\cdot10^7$ MD time steps). Note that $C_{AA}(\tau)$ is normalized in such a way that it's initial value at $\tau=0$ is always 1 (Eq. \ref{pearson_c}). Just like in the case of the PDF of the total area, we found that the auto-correlation function does not depend on the particular realization of the dynamics (a particular random sequence of excited springs).

Fig.~\ref{without_feedback}(c) shows the area auto-correlation functions for periodic and non-periodic excitations, respectively. In the case of periodic excitations, $C_{AA}(\tau)$ oscillates around a decreasing envelope and, instead of decaying to zero at longer delay time $\tau$, it undergoes steady oscillation between positive and negative values. In the case of non-periodic excitations, $C_{AA}(\tau)$ decays to zero at higher values of $\tau$ (a similar slow decay of the envelope of the oscillating correlation function is observed in the periodic case). In this case the auto-correlation function can be fitted by an exponential with the characteristic decay time of $6$ time units (this time also characterizes the decaying envelope of the oscillatory auto-correlation function in the case of periodic excitations). 

\subsection{Feedback through area-pressure coupling:}

In the previous section, we considered the case of constant hydrostatic pressure $p$. In the current section, we examine the case where $A$ and $p$ are coupled i.e., changes of the area induce changes of pressure (e.g., through actively driven osmosis) which again affect the total surface area, and so on. In the following we assume that the excitations are non-periodic (as discussed in the previous section), and consider both (i) negative and (ii) positive feedback. 
\par In the case of negative feedback, an increase (decrease) of area $A(t)$ results in decrease (increase) of hydrostatic pressure $p(t)$ as 
\begin{equation}
    p(t)=p_0\Big(1-\frac{A(t)-A_0}{A_{max}-A_0}\Big)
    \label{negative_feedback}
\end{equation}
where $p_0=0.1$, and $A_0=651.36$ is the area value at $t=0$ (the surface area of the system in steady state with pressure $p_0$ but without active excitations). $A_{max}$ is taken to be bigger than both $A(t)$ and $A_0$ such that the value of $p(t)$ is always positive.

\par In the case of positive feedback an increase (decrease) of area $A(t)$ results in increase (decrease) of osmotic pressure $p(t)$. The pressure and the area are related in this case as 
\begin{equation}
    p(t)=p_0\Big(1+\frac{A(t)-A_0}{A_{max}-A_0}\Big)
    \label{positive_feedback}
\end{equation}
 The value $A_{max}=1100$ has been chosen such that $p(t)$ is always positive, for both types of feedback.

\begin{figure}[t!]
        \centering
        \includegraphics[width=1.0\linewidth]{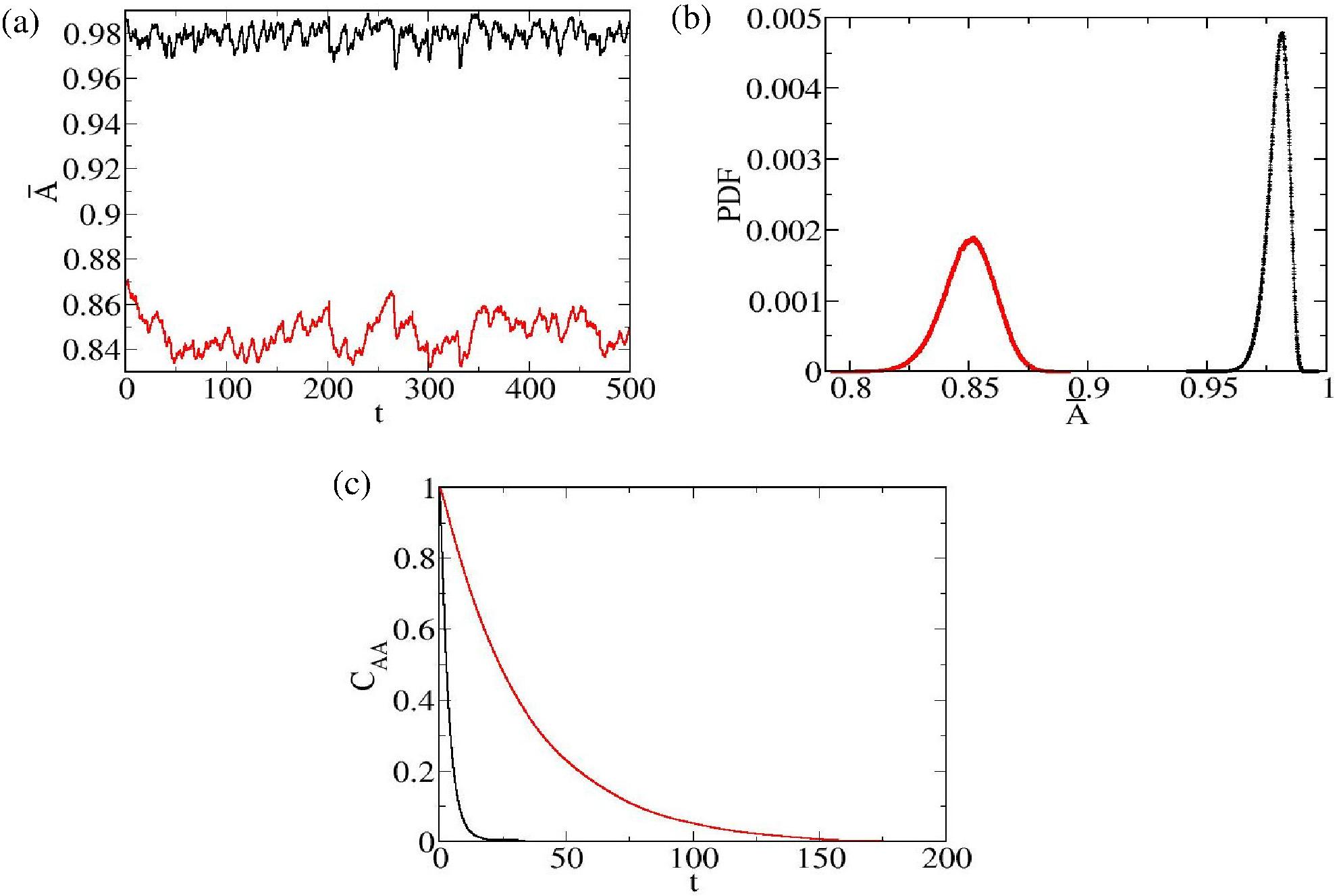}
        \caption{Comparison of (a) time series, (b) probability distribution function, (c) auto-correlation function of the total area, for negative (in black) and positive (in red) feedback. }
        \label{with_feedback}
\end{figure}

In Fig. \ref{with_feedback}(a) we present the time series of the surface area for  systems with negative and positive feedback. One observes that the mean area is dramatically reduced and that area fluctuations are much stronger and are correlated over longer time scales, in the presence of positive feedback compared to the negative feedback case. The statistical analysis of these area fluctuations is presented below.

In Fig. \ref{with_feedback}(b) we plot the probability distributions of normalized total surface area $\bar{A}$, with and without feedback (since the pressure $p$ defined in Eqs. \ref{negative_feedback} and \ref{positive_feedback} is a linear function of the area $A$, its distribution is determined by that of $A$ and is not shown here).
The area distributions for positive feedback (red curves) are broader than the corresponding distributions for negative feedback (black curves). This can be understood in the following way: in case of positive feedback, an increase of the total area increases the pressure which again increases the total area and so on. Conversely, a decrease in the total area of the system decreases the pressure which then decreases the total area, etc. This runaway mechanism broadens the area distribution. Conversely, when feedback is negative, an increase of total area results in decrease of pressure which then decreases the area and, in turn, increases the pressure, etc. (similarly to Le Chatelier principle which guarantees the stability in thermodynamics). This effect leads to narrowing down of the corresponding area  distribution. 

In Fig. \ref{with_feedback}(c) we plot the auto-correlation function of the area $C_{AA}(\tau)$ as a function of the delay time $\tau$, for non-periodic excitations with negative (black curve) and positive feedback (red curve). Both correlation functions decay exponentially with characteristic decay times $3.5$ and $33.4$ for negative and positive feedback cases, respectively. The more than an order of magnitude slower decay of correlations in the positive feedback case reflects the persistent nature of the corresponding dynamics as discussed in the preceding paragraph.

\subsection{Local properties: areas of triangles}
\begin{figure}[t!]
        \centering
        \includegraphics[width=1.0\linewidth]{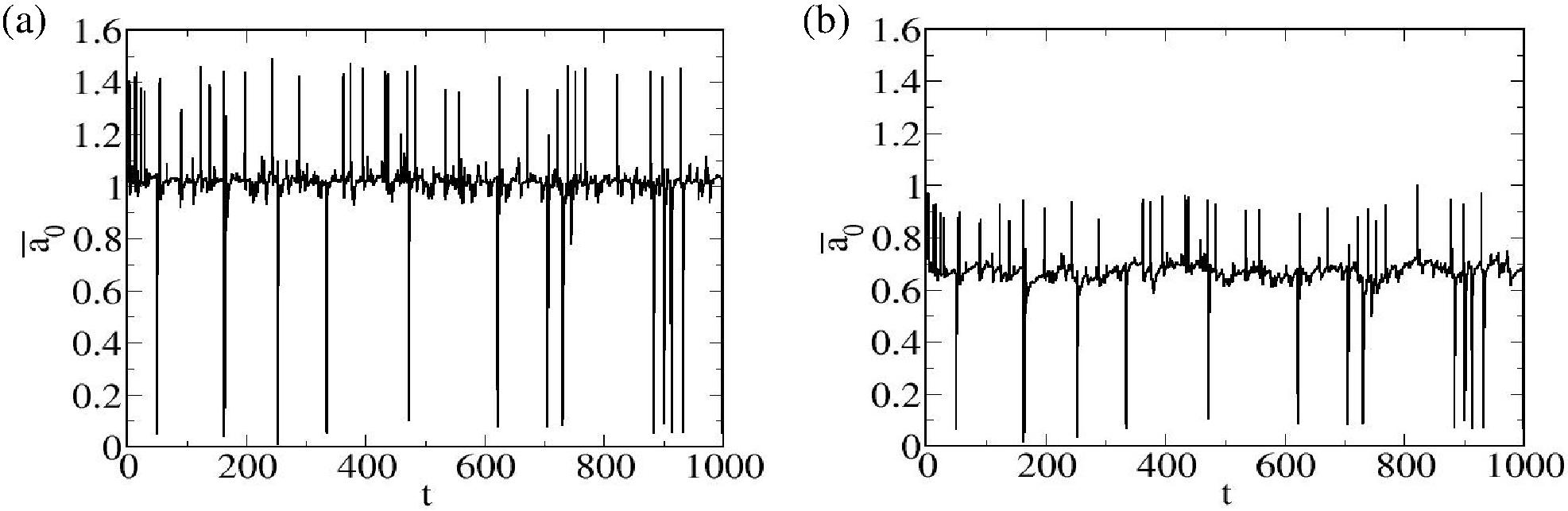}
        \caption{Plot of the dimensionless area ${\bar a}_0$ of triangle $0$ as a function of time, for (a) negative and (b) positive feedback cases.}
        \label{local_properties_1}
\end{figure}

Having studied the response of global properties (the total area) of the system to active perturbations, we now turn our attention to a local property: the deformation of individual triangles that play the role of surface elements in our discrete system. Since not all triangles  have identical sizes even in the ground state (i.e., with $p_0=0$), upon choosing a triangle whose temporal history we wish to follow and  labelling it by $0$, we normalize the area of this triangle by its steady state value in the absence of excitations $a_0^{s}$.
In Fig. ~\ref{local_properties_1} we plot ${\bar a}_0=a_0/a_0^{s}$ as a function of time, for both negative and positive feedback cases.

Inspection of Figs. ~\ref{local_properties_1}(a) and (b) shows that the mean (averaged over time) area of the triangle is lower in the positive than in the negative feedback case, in agreement with our previous analysis of the global response of the network. Note that since the same sequence of contractile excitations of springs in the network is used in both cases, the times at which the peaks and the dips in Figs. ~\ref{local_properties_1}(a) and (b) take place are identical but their amplitudes are not. We find that the deepest dips of ${\bar a}_0(t)$ that correspond to ${\bar a}_0(t)< 0.1$, occur when one of the springs of the chosen triangle is excited (not shown). No such simple identification can be made for the smaller peaks and dips observed in Figs. ~\ref{local_properties_1}, but since contraction of a spring is accompanied by stretching of its nearest neighbors and by contraction of its next nearest neighbors (see Fig.~\ref{single_excitation_length}(c)), we argue that the majority of these peaks and dips correspond to excitations of neighboring triangles. 

\begin{figure}[t!]
        \centering
        \includegraphics[width=0.8\linewidth]{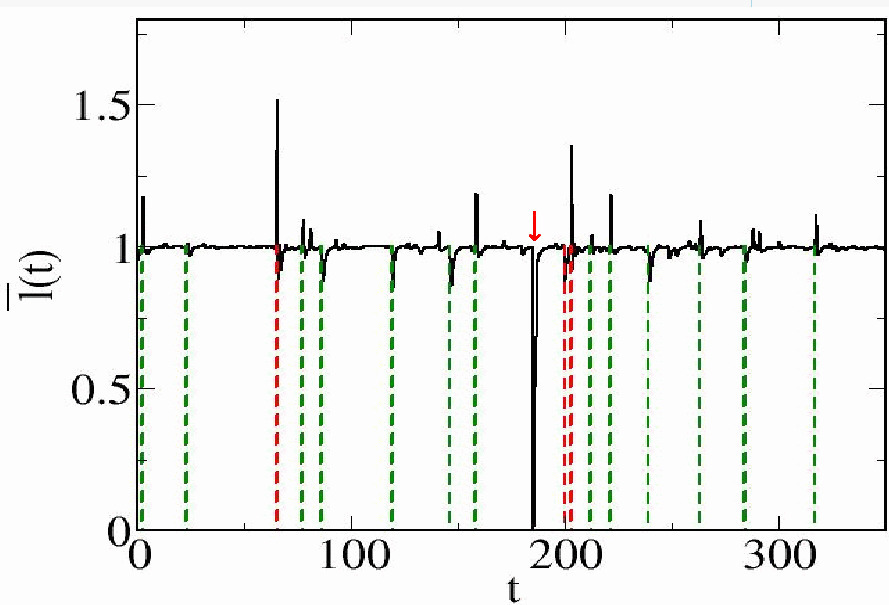}
        \caption{Time trace of the normalized elongation of a spring in a periodically excited network is shown in black. The red arrow indicates an excitation of the chosen spring. The broken vertical red and green lines correspond to excitations of nearest neighbor and next nearest neighbor springs, respectively.}
        \label{local}
\end{figure}

This identification is supported by Fig. \ref{local}, in which we present a time trace (shown in black) of the normalized length $\bar l$ of a particular spring in a periodically excited network, and denote by broken vertical red (green) lines events in which one of its nearest (next nearest) neighbors has been excited. Note that the locations of these lines coincide with the location of most of the peaks and dips in the time trace of the chosen spring (except for the dip denoted by a red arrow in Fig. \ref{local} that corresponds to excitation of this spring). This identification is not perfect since because of the slow relaxation of the network area, there is coupling between different excitations even in the periodic excitation case.

Additional information about the statistical properties of the local area dynamics in response to random excitations can be obtained from the distribution of dimensionless area of the triangle ${\bar a}_0$ (Fig. S6 in SI). In all cases the distributions have tails towards smaller triangle areas. This agrees with the observed asymmetry between extreme events (peaks and dips) in Fig. ~\ref{local_properties_1}. The peak of the distribution for negative feedback (black curve) occurs at a higher value of ${\bar a}_0$ than the peak of the distribution for positive feedback (red curve), in agreement with the observed difference in the positions of the baselines (compare Figs. \ref{local_properties_1}(a) and (b)). A similar effect was observed for the total area distributions in Fig. \ref{with_feedback}(b), and was attributed to the combined effect of active contractions and of the resulting pressure changes: the latter enhance (suppress) the contraction-induced reduction of the area in the positive (negative) feedback case. 
 
Next, we computed the cross-correlation function between the area ${\bar a}_{0}$ of  triangle $0$ and the areas ${\bar a}_{j}$ of five other triangles, whose centers of mass are located at different spatial distances $r_{0j}$ from the center of mass of triangle $0$, that range from $r_{01}=0.822$ (for neighboring triangles that share a common side) to $r_{05}=6.718$ (for the furthest triangles in the system).  The cross correlation function is defined as
\begin{equation}
    C_{ij}(\tau)=\frac{<{\bar a}_{j}(t+\tau){\bar a}_{i}(t)>-<{\bar a}_{i}(t)><{\bar a}_{j}(t)>}{\sqrt{(<{\bar a}_{j}(t)^2>-<{\bar a}_{j}(t)>^2)(<{\bar a}_{i}(t)^2>-<{\bar a}_{i}(t)>^2)}}
\end{equation}
 \begin{figure}[t!]
        \centering
        \includegraphics[width=0.8\linewidth]{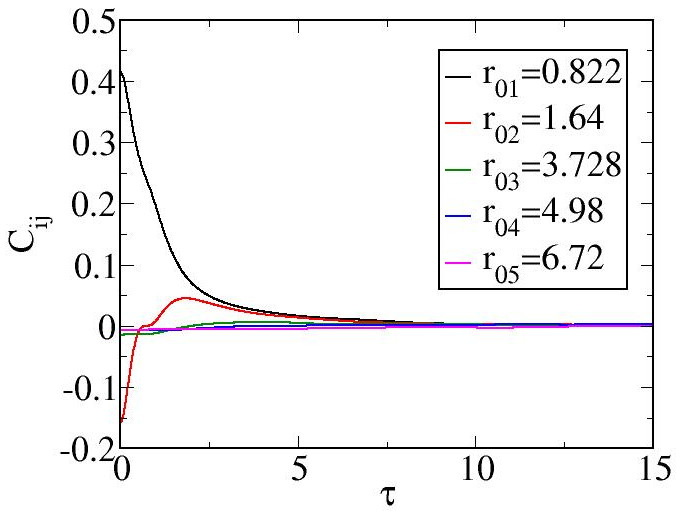}
        \caption{Cross-correlation between areas of different triangles as a function of the delay time $\tau$, for a network without area-pressure feedback.}
        \label{local_properties_3}
\end{figure}

The cross correlation functions $C_{0j}(\tau)$ for each of the  triangles $j=1-5$ are plotted as a function of the time shift $\tau$ in Fig. \ref{local_properties_3}, for the case of no feedback. While areas of triangles that share a common spring ($r_{01}$) are positively correlated, those that share a common vertex but not a common side ($r_{02}$) are slightly anti-correlated at short delay times. We attribute the latter effect to the fact that contractile excitation of a spring is followed by stretching of its nearest neighbors (see Fig. \ref{single_excitation_length}(c)).  The correlations between more distant triangles are negligible at all times. 

One may wonder whether the small anti-correlation between triangles that share a common vertex (see Fig. \ref{local_properties_3}) is significant. Note that since the correlation functions have been obtained by averaging the products ${\bar a}_{j}(t+\tau){\bar a}_{0}(t)$ over time $t$, they are dominated by typical fluctuations about the mean and not by rare extreme events associated with large dips and peaks in Figs. ~\ref{local_properties_1}(a) and (b). In order to focus on the contribution of these rare events, in Fig. S7(a) in SI we present a scatter plot of the areas of two neighboring triangles that share a common vertex (each point corresponds to a pair of values, $\bar a_0$ and $\bar a_2$, measured at time $t$). An attractor that exhibits the negative correlation between the two neighboring triangles is clearly observed in this figure. As expected from Fig. \ref{local_properties_3}, no such correlations are observed in the scatter plot of distant triangles, $\bar a_0$ and $\bar a_5$ (see Fig. S7(b) in SI).

\section{Discussion:}
Using a triangulation of a closed surface, we constructed a simple network of nonlinear springs to model a thin elastic shell swollen by hydrostatic pressure. We  carried out computer simulations to study the dynamics of this network, in response to active (non-thermal) excitations produced by contraction and stiffening of randomly chosen springs. We computed the statistical properties of the resulting fluctuations of the global (total surface area) and local (surface elements) attributes of the system. This was done both for constant hydrostatic pressure and in the presence of feedback, when the pressure can readjust to fluctuations of the surface area. 

In the constant pressure case we found that the peaks of the distributions of the total area are shifted downwards compared to their steady state values (without excitations), consistent with the contractile character of the excitations. Both the shift of the peak and the width of the distribution (the magnitude of fluctuations) are much stronger in the non-periodic than in the periodic case. The physical reason behind these observation is that while in the periodic case several excitations can exist simultaneously in the network, only one excitation is present at any instant of time in the periodic case.

Additional information was obtained from the auto-correlation functions of the total area. While for periodic excitations, oscillations between positive and negative correlations persist for arbitrarily long delay times, simple exponential decay of correlations is observed for non-periodic excitations. The characteristic decay time is much slower than the microscopic time scales (the relaxation time of the excitations and the average time interval between successive excitations) and reflects the much slower relaxation dynamics of the entire network that gives rise to the low frequency, large amplitude fluctuations observed in the time trace of the total area. These global relaxation modes depend on the hydrostatic pressure - increasing the pressure increases both the amplitude and the frequency of slow  fluctuations of the area of the system (see Figs. S3(a) and (b) in SI).

We explored the effects of coupling between surface area and pressure: when feedback is positive, increase of total area increases pressure which, in turn increases the area, and so on. This creates persistence and amplifies fluctuations: slow large amplitude  fluctuations of the area maintain their direction (periods in which the area increases/decreases) for much longer periods of time.  Such persistent effects give rise to the observed broad area distributions and slow decay of the area auto-correlation function.

Next we studied the dynamics of local properties of the system, namely the area fluctuations of individual triangles that play the role of surface elements in our discrete system. The time trace of area fluctuations of a triangle contains strong dips (it's area falls below $10\%$ of the steady state value) each of which corresponds to active excitation of a spring that forms one of the sides of the triangle. Other observed peaks and dips are attributed to excitations of springs that belong to neighboring triangles. As evident from the cross-correlation function of the areas of different triangles, at equal times both correlations and anti-correlations exist for neighboring triangles that share a side or a vertex, respectively, but these correlations decrease with increasing separation between triangles and with delay time.  We therefore conclude that active excitations generate strong localized perturbations in the network which decay rapidly with time and with distance from the source of excitation. 

Since our results were obtained for a particular network topology (that of a triangulated fullerene) one may wonder about their generality. We believe that the emerging physical picture is applicable a broad class of active elastic shells swollen by hydrostatic pressure. The pressure  stretches the network and introduces a coupling between its global and local dynamics. Both the amplitude and the relaxation of the deformation created by a local excitation are controlled not only by its intrinsic parameters (the change of equilibrium length and spring constant and its intrinsic relaxation time),  but also by forces exerted on it by the rest of the network. Conversely, each local excitation produces a global deformation of the network as manifested by the shrinking of its surface area. For large enough systems (as long as both the excitation lifetime and the time between excitations are much shorter than the relaxation time of the total area), there is separation of time scales between the microscopic relaxation time of a surface element (a spring or a triangle) and that of the entire network. This guarantees that multiple excitation events will conspire to deform the network and strong fluctuations of its surface area will result. Such fluctuations of the initial spheroidal shape of developing hydra have indeed been observed in 
recent experiments \cite {braun2021calcium}. The modeling of other phenomena such as the transition from a spheroidal to a cylindrical shape during the process of development of the hydra, would probably require modification of the present model, in order to account for the intrinsic anisotropy of the actomyosin skeleton and for the possibility of non-local excitations of the network. These directions will be explored in a future study.

\section*{Conflicts of interest}
There are no conflicts to declare.

\section*{Acknowledgements}
We would like to acknowledge helpful discussions with Erez Braun and Orit Peleg. This work was supported by a grant from the Israel Science Foundation.



\section*{Supplementary material}

# Supplementary Information

# Network Model of Active Fluctuations of Thin Elastic Shells

Ajoy Maji and Yitzhak Rabin

Department of Physics and Institute of Nanotechnology and Advanced Materials, Bar-Ilan University, Ramat-Gan 5290002, Israel

# Distribution of spring lengths and triangle areas:

The equilibrium lengths of the 270 springs in the network are taken to be the lengths of the sides of the triangles in the triangulated fullerene and, therefore, are not same. The springs can be divided into four groups, each having a different value of equilibrium length $l_{eq}$ (Fig.1). Similarly, since the equilibrium lengths of all springs are not the same, the areas of all triangles are also not identical and can be divided into three groups, each having a different value of area $a_{eq}$.

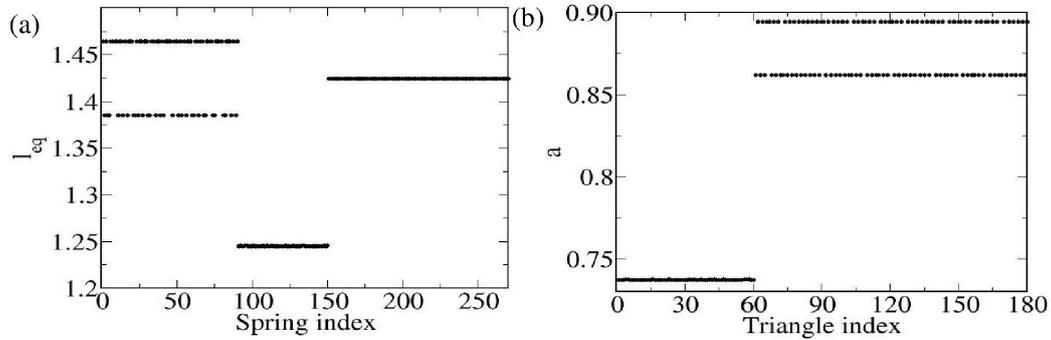

Supplementary Fig. 1: (a) Distribution of equilibrium lengths of different springs. (b) Distribution of area of each of the triangles.

# Integration algorithm

$$\frac{d}{dt}(e^{\zeta t}\, \vec{v}_k) = e^{\zeta t} \vec{F}_k$$

Integrating both sides of this equation,

$$\int_{-\Delta/2}^{t} dt' \frac{d}{dt'}(e^{\zeta t'}\, \vec{v}_k) = \int_{-\Delta/2}^{t} dt'\, \vec{F}_k\, e^{\zeta t'} \qquad (1)$$

where $\Delta$ is the MD time step which we take to be much shorter than the viscous damping time (mass/(friction coefficient)).

Solving equation 1, we get the following expressions for updating the velocity and position of vertex $K$:

$$\vec{v}_k((n+\frac{1}{2})\Delta) = \vec{v}_k((n-\frac{1}{2})\Delta)e^{-\zeta\Delta} + e^{-\zeta\frac{\Delta}{2}}\, \Delta\, \vec{F}_k(\vec{r}_k(n\Delta)) \qquad (2)$$

$$\vec{v}_k(t) = \vec{v}_k(-\frac{\Delta}{2})e^{-\zeta(t+\Delta/2)} + e^{-\zeta t}\int_{-\Delta/2}^{t} dt' \vec{F}_k(t')e^{\zeta t'}$$

$$\implies \vec{v}_k(\tfrac{\Delta}{2}) = \vec{v}_k(-\tfrac{\Delta}{2})e^{-\zeta\Delta} + \Delta \vec{F}_k(\vec{r}_k(t=0))e^{-\zeta\tfrac{\Delta}{2}}$$

In general,

$$\vec{v}_k((n+\frac{1}{2})\Delta) = \vec{v}_k((n-\frac{1}{2})\Delta)e^{-\zeta\Delta} + e^{-\zeta\frac{\Delta}{2}} \Delta \vec{F}_k(\vec{r}_k(n\Delta)) \qquad (3)$$

where $n$ is the number of MD time steps.

The equation for the position can be written as,

$$\vec{r}_k(t) = \vec{r}_k(0) + \int_0^t \vec{v}_k(t')dt'$$

At $t = \Delta$,

$$\vec{r}_k(\Delta) = \vec{r}_k(0) + \Delta\ \vec{v}_k(\Delta/2)$$

In other words,

$$\vec{r}_k((n+1)\Delta) = \vec{r}_k(n\Delta) + \Delta\ \vec{v}_k\left((n+\frac{1}{2})\Delta\right) \qquad (4)$$

We are using equation 3 and 4 for updating velocities and positions of the vertices

## Relaxation to steady state:

Beginning with a network in equilibrium in the absence of hydrostatic pressure, at time, $t = 0$, we suddenly switch on hydrostatic pressure by changing $p_0$ from 0 to 0.1. With the simulation parameters used in this paper, it takes about $40 - 50$ time units to reach a steady state characterized by a new pleatea value of the total surface area $A$ (Fig. 2). Increasing $p_0$ from 0 to a higher value of pressure, increases the steady state value of the surface area but decreases the relaxation time compared to the values of shown in Fig. 2.

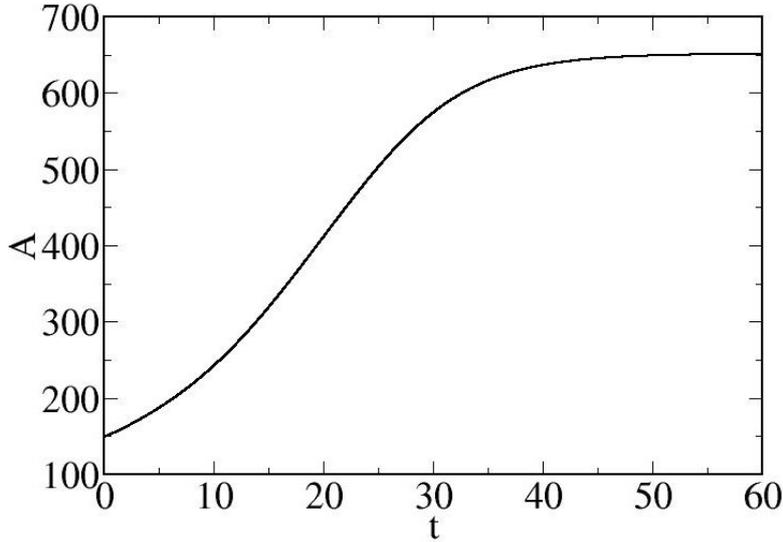

Supplementary Fig. 2: (a) The total surface area $A(t)$ of the network plotted as a function of time following a step-like increase of hydrostatic pressure from $p_0 = 0$ to $0.1$.

# Single excitation dynamics: effect of hydrostatic pressure:

The effect of hydrostatic pressure on excitation and relaxation dynamics of a single spring and total surface area of the system can be observed by comparing the steady state results for $p_0 = 0.05$ and $0.1$. Since in steady state with $p_0 = 0.1$ the network is more inflated than for $p_0 = 0.05$, the elastic energy stored in the springs is higher as well in case of $p_0 = 0.1$. Once a single spring is excited, $\bar{l}$ reaches its minimum value faster and the maximal contraction of the spring is higher in the $p_0 = 0.1$ than in the $p_0 = 0.05$ case. Since the deformation of the spring is opposed by network forces that increase with the inflation of the network, the excited spring relaxes faster to its steady state value in the $p_0 = 0.1$ than in the $p_0 = 0.05$ case (Fig. 3(a)). (b) Faster excitation and relaxation dynamics at higher hydrostatic pressure can also be observed in the normalized total surface area $\bar{A}$ versus time plot (Fig. 3(b)).

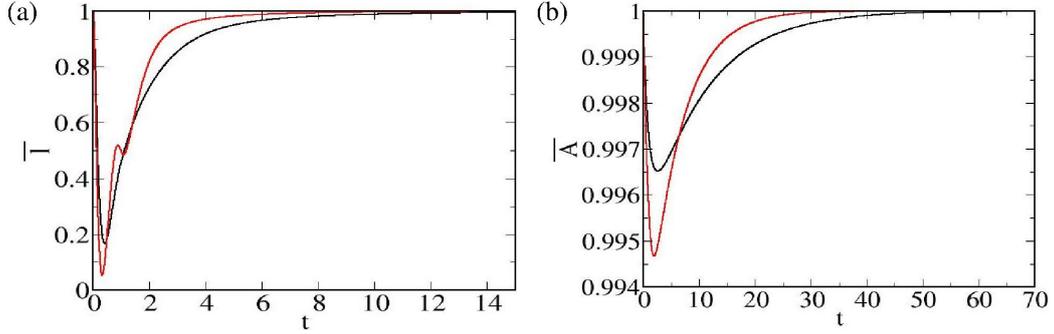

Supplementary Fig. 3: (a) The normalized length $\bar{l}$ of an excited spring and (b) the normalized total area $\bar{A}$ of the network, are plotted as a function of time following the excitation, for $p_0 = 0.05$ (black curve) and $0.1$ (red curve). The relaxation time of the spring is $t_r = 1$.

# Single excitation dynamics: Effect of spring constant $K$ and mass $m$

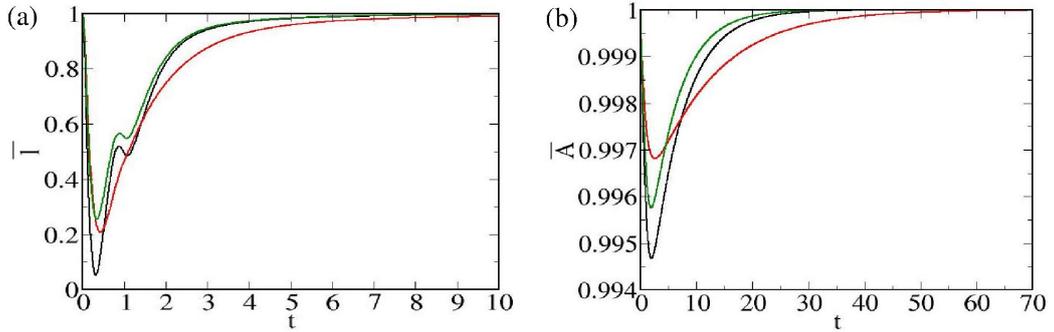

Supplementary Fig. 4: (a) The normalized length $\bar{l}$ of an excited spring and (b) the normalized total area ($\bar{A}$) as a function of time for different combinations of spring constants $K$ and mass $m$: $K = 1, m = 1$ (black); $K = 1, m = 2$ (red) and K=2, m=1 (green).

# Distribution of number of excited bonds during non-periodic excitations:

During non-periodic excitations, the time interval between two consecutive excitations is chosen from an uniform distribution with a mean of 1 time

5

unit. In this case the number of excitations simultaneously present in the network, fluctuates in time. In Fig. 4 we plot the probability distribution of the number of simultaneous excitations (events were accumulated during the measurement time). Both the peak and the mean of the distribution occur at 1, corresponding to a single excitation).

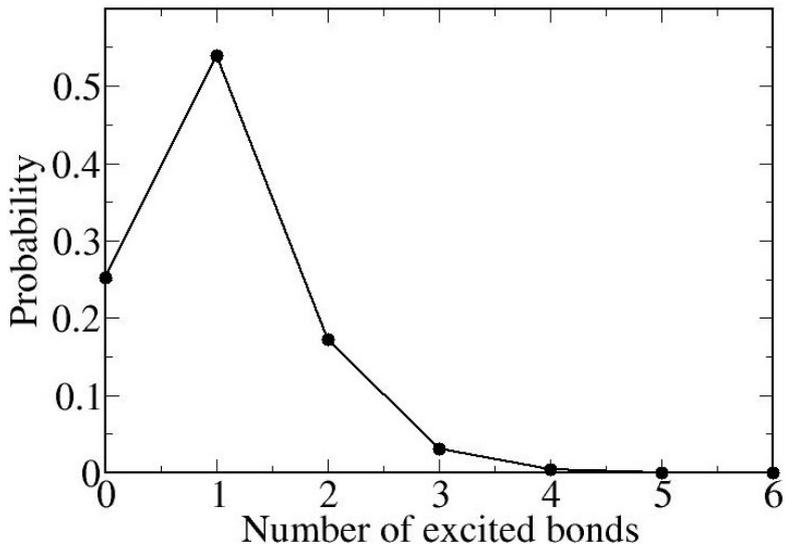

Supplementary Fig. 5: Distribution of number of excited springs present in the system at a particular instant of time during non-periodic excitations without A-p feedback.

## Distribution of normalized areas of triangles:

In Supplementary Fig. 6 we plot the PDFs of normalized area of triangles ($\bar{a}$) for the case of non-periodic excitations, with no total area-pressure feedback and with negative and positive feedback. The three distributions have similar widths and tails towards lower values of $\bar{a}$. The peak of the positive feedback distribution is shifted to lower area compared to the other two.

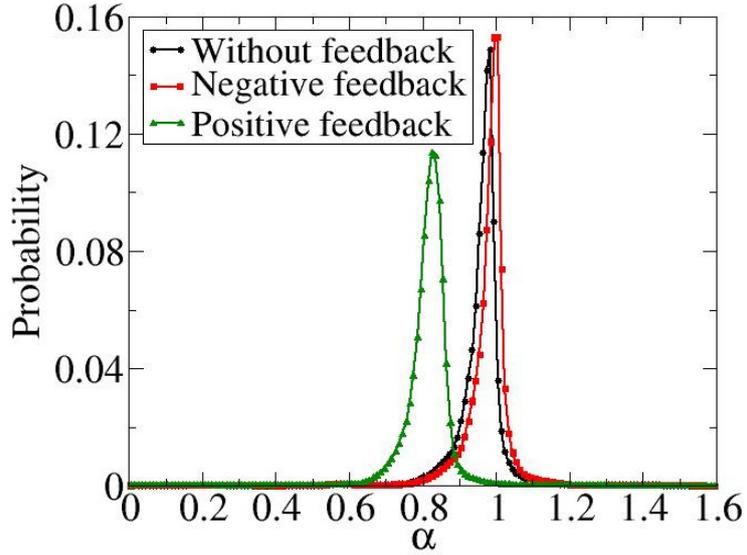

Supplementary Fig. 6: Distribution of normalized area of triangles ($\bar{a}$) in case of without feedback (black), negative feedback (red) and positive feedback (green).

## Scatter plots of areas of triangles:

In Fig. 7 we present scatter plots of the instantaneous values of areas of two (a) neighboring and (b) distant triangles. While there are nearly no correlation between distant triangles (see Fig. 7(b)), negative correlations are observed between triangles that share a common vertex, in Fig. 7(a)

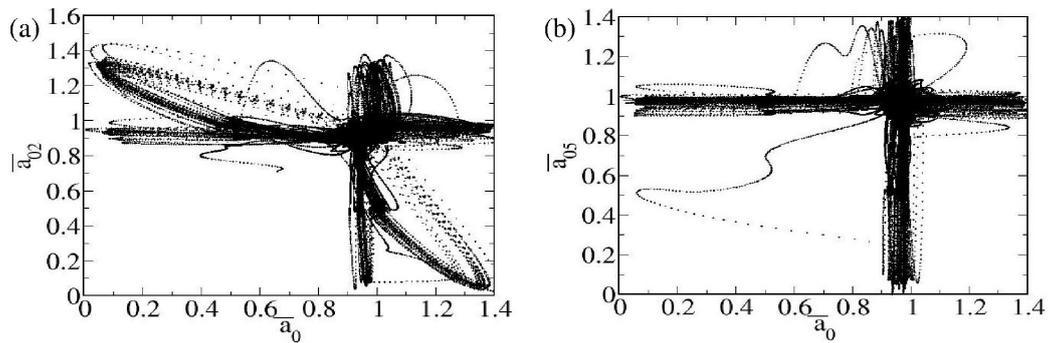

Supplementary Fig. 7: Scatter plots of areas of two neighboring (a) and distant (b) triangles.

# Supplementary movie 1

The movie shows the active excitations in the network when the excitations are introduced non-periodically in the network.



\end{document}